\documentclass[10pt]{iopart}
\usepackage{graphicx}
\graphicspath{{./Figures/},}
\usepackage{bm}
\usepackage{multirow}
\usepackage[table]{xcolor}
\usepackage[export]{adjustbox}
\makeatletter
\let\old@makecaption=\@makecaption
\usepackage{subcaption}
\let\@makecaption=\old@makecaption
\makeatother
\usepackage{color}
\usepackage{orcidlink}
\newcommand{\RuO}{RuO$_{2}$ }

\begin{document}

\title{Crystal structure and absence of magnetic order in single-crystalline RuO$_2$}

\author{L. Kiefer$^1$\orcidlink{0009-0008-5716-2816}, F. Wirth$^1$\orcidlink{0000-0002-6386-029X}, A. Bertin$^1$\orcidlink{0000-0001-5789-3178}, P. Becker$^2$\orcidlink{0000-0003-4784-3729}, 
	L.~Bohat\'{y}$^2$\orcidlink{0000-0002-9565-8950}, K. Schmalzl$^3$\orcidlink{0000-0003-4836-5642},  A. Stunault$^4$,   J. A. Rodr\'iguez-Velamazan$^4$,   O. Fabelo$^4$, M.~Braden$^1$\orcidlink{0000-0002-9284-6585}}
\address{$^1$Institute of Physics II, University of Cologne, 50937 Cologne, Germany}
\address{$^2$Sect. Crystallography, Institute of Geology and Mineralogy, University of Cologne, 50674 Cologne, Germany}
\address{$^3$ Forschungszentrum Jülich GmbH, Jülich Centre for Neutron Scienc at ILL, 71 avenue des Martyrs, 38000 Grenoble, France}
\address{$^4$Institut Laue-Langevin, 71 avenue des Martyrs, 38042 Grenoble, France}
\mailto{kiefer@ph2.uni-koeln.de}
\mailto{braden@ph2.uni-koeln.de}

\begin{abstract}
	
RuO$_2$ was considered for a long time to be a paramagnetic metal with an ideal rutile-type structure down to low temperatures, but recent studies on single-crystals claimed evidence for antiferromagnetic order and some symmetry breaking in the crystal structure. We have grown single-crystals of RuO$_2$  by vapor transport using either O$_2$ or TeCl$_4$ as transport medium. 
These crystals exhibit metallic behavior following a $T^2$ low-temperature relation and a small paramagnetic susceptibility that can be attributed to Pauli paramagnetism. 
Neither the conductance nor the susceptibility measurements yield any evidence for a magnetic or a structural transition between 300\,K and $\sim$4\,K. Comprehensive single-crystal diffraction studies with neutron and X-ray radiation reveal the rutile structure
to persist until 2\,K in our crystals, and show nearly perfect stoichiometry. 
Previous observations of symmetry forbidden reflections can be attributed to multiple diffraction. Polarized single-crystal neutron diffraction experiments at 1.6\,K
exclude the proposed antiferromagnetic structures with ordered moments larger than 0.01 Bohr magnetons.
	
\end{abstract}
\vspace{2pc}
\noindent{\it Keywords\/}: Altermagnetism, Crystal structure, Polarized neutron diffraction, RuO$_2$

\submitto{\JPCM}
\maketitle
\ioptwocol

\section{Introduction}

Altermagnetism represents a novel phase of magnetism, characterized by a combination of ferromagnetic and antiferromagnetic properties, which gives rise to a wide range of fascinating phenomena \cite{Smejkal.2022,Smejkal.2022b}. Typically, collinear magnets are classified into two distinct categories: ferromagnetic and antiferromagnetic systems. Ferromagnetic systems posses finite magnetisation, whereas antiferromagnetic systems exhibit a compensated magnetisation. While ferromagnets exhibit a uniform spin splitting of electronic bands, the bands in most antiferromagnets remain spin degenerate, because the up and down spin sites are related by translation or inversion symmetry. Smejkal $et$ $al.$ \cite{Smejkal.2022b} discovered that the situation is distinct when the antiparallel spin sublattices are related by a different symmetry operation, such as a rotation. In this case, spin splitting of bands emerges despite the presence of fully compensated magnetisation. These materials are named altermagnets \cite{Smejkal.2022b,Smejikal.2023} referring to the alternating arrangement of both the spin moments and of the ligand surroundings.
The combination of spin splitting with a linear magnon dispersion relation has generated an enormous interest in this new class of materials, as it allows well defined magnon pulses and numerous potential applications in spintronics and magnonics \cite{Smejikal.2023}.\\
\begin{figure*}
	\centering
	\includegraphics[scale=0.4]{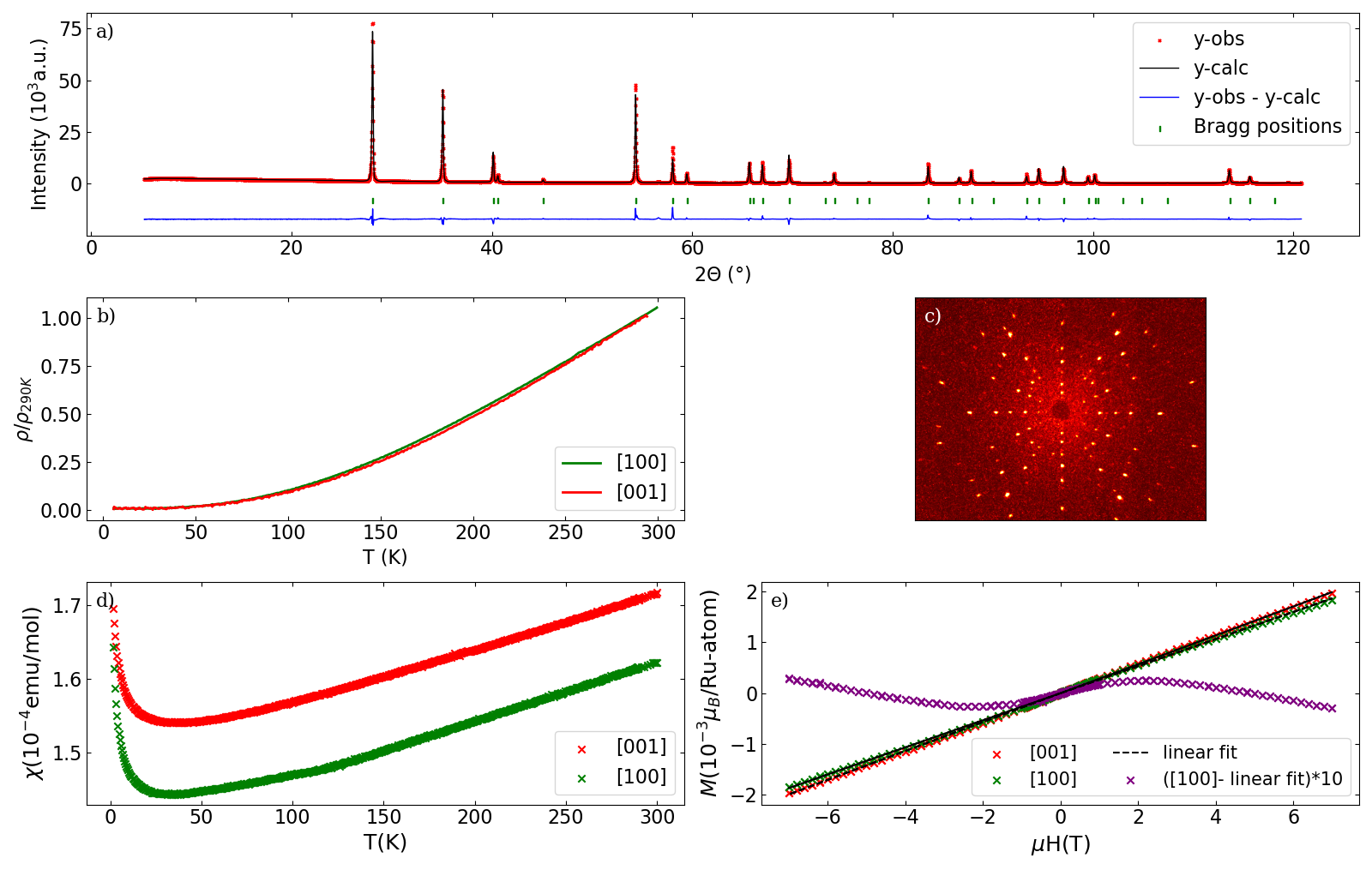}
	\caption{Characterisation of the \RuO single crystals. a) Small crystals were crushed to obtain a powder sample of RuO$_2$. X-ray diffraction data obtained at room temperature reveals no impurity peaks and Rietveld analysis gives a R\textsubscript{wp}-value of 8.15\,\%. b) Resistivity measured along and perpendicular to the tetragonal axis; $\rho_{290K}$=$124\, \Omega cm$. c) Neutron Laue images taken on $OrientExpress$ at the ILL with the incoming beam parallel to [100]. d) Temperature dependence of magnetic susceptibility measured along [100] and [001] directions with a field of $0.5\, T$. e) Magnetisation along [100] and [001] directions for magnetic fields up to 7\, T at 2\, K. The mangenta crosses in e) present the difference between the data and the linear fit multiplied by a factor 10.}
	\label{Charac}
\end{figure*}

There are several known materials which realize altermagnetic order \cite{Smejkal.2022b,Reimers.2024,Aoyama.2024,Krempasky.2024}, but in view of spintronics applications metallic systems are most interesting \cite{Park.2011}. Therefore, metallic \RuO \ has been identified as a prototype material \cite{Smejkal.2022b}. The predicted existence of the anomalous Hall effect \cite{Feng.2022} and the spin-to-charge conversion efficiency \cite{Bai.2023}  as well as the spin splitting \cite{Lin,Fedchenko} have been confirmed for RuO$_2$. However, most of the latter experiments were performed on film samples.
 
\RuO crystallizes in the rutile structure, spacegroup $P4_2/mnm$, and exhibits good metal properties, with room temperature resistivity values of only $50\, \mu\Omega$cm and large residual resistivity ratios \cite{Bolzan.1997,Schafer.1963,Ryden.1970}. Berlijn $et$ $al.$ \cite{Berlijn.2017} reported the observation of nuclear Bragg peaks violating the non-symmorphic symmetry elements $4_2$ and $n$, but the underlying structural distortion could not be determined. In contrast no evidence for a structural phase transition was found in the electrical transport \cite{Lin.2004,Glassford.1994}, heat capacity \cite{ONeill.1997,Cordfunke.1989} or thermal expansion \cite{Touloukian.1972} measurements. \RuO \ bears strong application potential as catalyst
and in  microelectronics \cite{Over.2012} due to its combination of a high electrical conductivity with excellent thermal and chemical stability \cite{Torun.2013}. 

Regarding the magnetic properties,  \RuO \ was assumed to be a Pauli paramagnet \cite{Ryden.1970,Over.2012,Mukuda.1999} for a long time. However, recently antiferromagnetic order occurring even above room temperature was claimed by polarised neutron scattering \cite{Berlijn.2017}, although a very small ordered
moment of only 0.05$\mu_B$ contrasts with the assumed high transition temperature.
The antiferromagnetic order in \RuO \ was confirmed using resonant X-ray diffraction \cite{Zhu.2019}. 
In the proposed antiferromagnetic structure model, the up and down spins are connected through the 
non-symmorphic symmetry operations $4_2$ or $n$ rendering RuO$_2$ a simple metallic altermagnet.
However, the magnetic order became a topic of active discussion, when recent $\mu$SR studies 
did not observe any evidence for magnetic order \cite{Hiraishi.2024}. Due to the high sensitivity of the muon, ordered moments
an order of magnitude smaller than those reported by Berlijn $et$ $al.$ \cite{Berlijn.2017} would have been observed \cite{Hiraishi.2024}. This observation was confirmed via another $\mu$SR experiment  \cite{Keler.2024} combined with neutron diffraction experiments that also could not detect magnetic order. 
Furthermore, an earlier nuclear magnetic resonance study finds the nuclear relaxation time to be about two orders of magnitude smaller than those in CaRuO$_3$ or Sr$_2$RuO$_4$ concluding a non-magnetic character similar to Ru metal \cite{Mukuda.1999}.
The occurrence of magnetic order in stoichiometric \RuO \ was further questioned by density functional theory calculations that require an unrealistic value of the correlation strength $U$ to stabilize magnetic order \cite{Smolyanyuk.2024b}. These calculations propose that magnetic order can arise from doping through
a moderate amount of Ru vacancies \cite{Smolyanyuk.2024b}.

In summary, there are three open questions, namely whether there is a structural distortion in bulk RuO$_2$, whether stoichiometric \RuO \ exhibits magnetic order, and whether such magnetic order in \RuO is affected by vacancies.
Due to the importance of \RuO for the ongoing discussion about altermagnetism we have performed X-ray, unpolarized neutron and polarized neutron diffraction experiments on well-characterised crystals.  We find no structural distortion, which would lower the symmetry of RuO$_2$, and we can exclude the reported magnetic order in \RuO for moments about ten times smaller than the reported ones.\\

\section{Experimental}

For the growth of large single crystals of \RuO two well-developed methods of chemical transport reaction (CTR) growth that are based on either higher Ru oxides (RuO$_3$, RuO$_4$)\cite{Schafer.1963,Shafer_1979,Huang.1982} or on RuCl$_4$ \cite{Oppermann} as volatile species of the transport reaction are reported in literature. The former method uses growth in a O$_2$ flow while for the later method TeCl$_4$ is added as transporting agent in closed ampoules. In both cases Ru or \RuO can be applied as educt. Using CTR growth in flowing O$_2$ we obtained large single crystals of \RuO of serveral millimeters size, but also growth in closed ampoules using TeCl$_4$, yielded millimetersize crystals \cite{NaglerBohatyPentinghaus1991}. In the following we indicate crystals grown in flowing O$_2$ with RuO$_2$-(O$_2$) and crystals grown with TeCl$_4$ with RuO$_2$-(TeCl$_4$).
We crushed some crystals to a powder sample in order to perform
powder X-ray diffaction experiments in Bragg-Brentano geometry on a $Stoe$ diffractometer using Cu K$_{\alpha}$ radiation. 
The data were analyzed by the Rietveld method with the Fullprof software package \cite{Rodriguez.2001}. 
The resistivity measurements were carried out with a standard four-probe method
by cooling the sample with liquid He. Magnetic properties were measured using a quantum design MPMS-XL7 superconducting quantum interference device magnetometer. 
\begin{table*}
	\caption{Refined structural parameters for different temperatures using X-ray and neutron single-crystal diffraction data.}
	\begin{indented}
		\item[]\begin{tabular}{@{}llllll}
			\br
			sample & \multicolumn{2}{l}{RuO$_2$-(O$_2$)} &\multicolumn{2}{@{}l}{RuO$_2$-(TeCl$_4$)} & RuO$_2$-(O$_2$) \\
			\mr
			method& X-ray & X-ray & X-ray & X-ray &neutron\\
			T(K)& 250& 80&250&80&2\\
			N& 18639&19853 & 15976& 18657&883\\
			unique N&227 &176 &232 &287 &164\\
			R (\%) &1.14 &1.3 &1.85 &2.14 &2.65\\
			wR (\%)& 2.49&2.35&4.65 &4.49 &3.76\\
			$\chi^2$& 2.05&1.85 &3.69 &3.35 &3.4665\\
			a(\AA)&4.4912(2) &4.4879(3) &4.4942(3) &4.4903(2) &4.4872(2)\\
			c(\AA)&3.1080(1) &3.1101(2) &3.1111(2) &3.1107(1) &3.1073(2)\\
			\mr
			Ru1 (0.5,0.5,0.5)&Occ=1.011(15) &Occ=1.039(12) &Occ=1.018(17) &Occ=1.013(13) &Occ=0.974(4)\\
			&U$_{\textrm{iso}}$=0.00243(13) &U$_{\textrm{iso}}$=0.00160(12) &U$_{\textrm{iso}}$=0.0025(2) &U$_{\textrm{iso}}$=0.0015(1) &U$_{11}$=0.0015(2)\\
			& & & & &U$_{33}$=0.0016(2)\\
			& & & & &U$_{12}$=0.00019(8)\\
			\mr
			O1 (x,x,0)&Occ=1 &Occ=1 &Occ=1 &Occ=1 &Occ=1\\
			&x=0.3056(2) &x=0.3057(2) &x=0.3059(1) &x=0.3062(1) &x=0.30603(6)\\
			&U$_{11}$=0.0037(2) &U$_{11}$=0.0014(3) &U$_{11}$=0.0036(2) &U$_{11}$=0.0021(2) &U$_{11}$=0.0037(2)\\
			&U$_{33}$=0.0028(3) &U$_{33}$=0.0016(4) &U$_{33}$=0.0029(3) &U$_{33}$=0.0024(2) &U$_{33}$=0.0029(2)\\
			&U$_{12}$=$-$0.0008(3) &U$_{12}$=$-$0.0004(3) & U$_{12}$=$-$0.0014(2)&U$_{12}$=$-$0.0006(2) &U$_{12}$=$-$0.00041(8)\\
			\br
		\end{tabular}
	\end{indented}
	\label{refinement}
\end{table*}

Comprehensive single-crystal X-ray diffraction experiments were conducted on a $Bruker$ D8 venture four-circle diffractometer with MoK$_{\alpha}$ radiation ($\lambda$ = 0.71 \AA) at temperatures of 80\,K and 250\,K. Reflection intensities have been integrated using $Apex4$ software and $4/mmm$ as Laue group, and absorption correction and scaling performed with the Multiscale algorithm. Refinements have been carried out with $Jana2020$ \cite{LukasPalatinus.0005} applying an extinction correction with an isotropic Becker-Coppens formalism \cite{Becker:a10603}.
Two large single crystals with a size of a few mm$^3$  were chosen for a diffraction study on the hot neutron four-circle diffractometer D9 at Institut Laue Langevin (ILL) in Grenoble ($\lambda$ = 0.835\,\AA).
These crystals were mounted with a [010] direction almost parallel to the $\phi$ axis.
The integrated intensities obtained on D9 were analyzed by structural refinements using Jana2020 \cite{LukasPalatinus.0005}. Neutron diffraction experiments exploiting polarisation analysis were performed on three crystals with the cold triple-axis spectrometer IN12 with an orange cryostat and a set of Helmholtz coils to determine the neutron polarization at the sample position (longitudinal polarization analysis). 
The incoming beam was polarized by a supermirror cavity and monochromatized with a (0,0,2) reflection
of pyrolithic graphite while we used a Heusler crystal for energy and polarization analysis in the outgoing beam \cite{Schmalzl.2016}. The flipping
ratio measured on the (2,0,0) nuclear Bragg reflection amounted to 21.5. Further polarized neutron diffraction experiments were realized on the hot neutron diffractometer D3 ($\lambda$ = 0.827\,\AA) with the cryopad device for spherical polarization analysis. On D3 we used a Heusler Cu$_2$MnAl (111) monochromator to polarize the incoming beam and a $^3$He spin filter to analyze the diffracted beam. Data obtained at IN12, at D3 and at D9 are available in Refs. \cite{KIEFERLara.2024,KIEFERLara.2024b,KIEFERLara.2024c}, respectively.

\begin{figure*}
	\centering
	\includegraphics[width=0.65\textwidth]{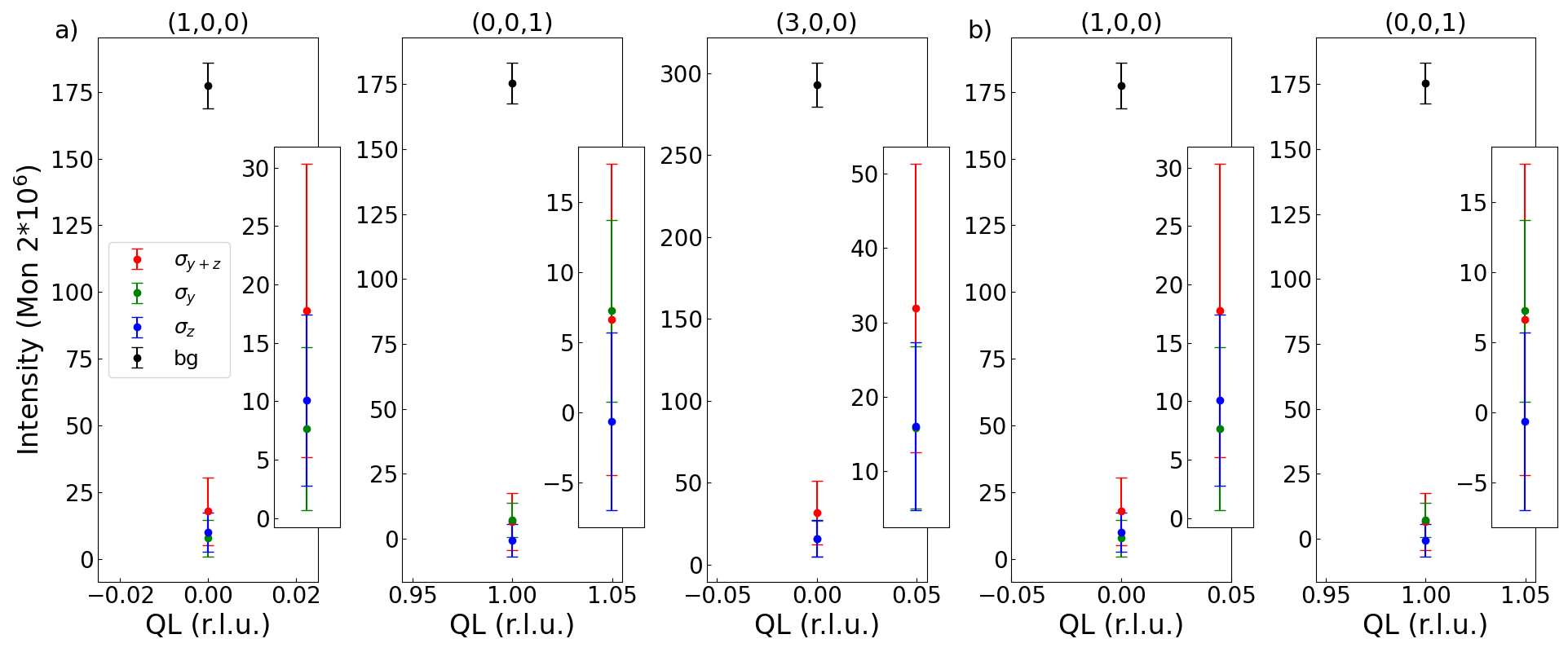}
	\caption{Polarization analysis of the possible magnetic peaks a) of RuO$_2$-(O$_2$) crystals and b) RuO$_2$-(TeCl$_4$) crystals  determined on IN12. Note that the (2,0,0) nuclear Bragg peak has an intensity of 350000\,cts with a monitor of 2$\times$ 10$^6$. The monitor is set to 2$\times$ 10$^6$\,cts correspondingto a counting time of $\sim$1530\,s at k$_i$=2.7\,\AA$^{-1}$ and $\sim$719\,s at k$_i$=1.85\,\AA$^{-1}$. Evaluation of the background is performed by $bg=I_{xsf}-I_{ysf}-I_{zsf}$ with I the intensity of the different channels.}
	\label{Polarisation_analysis}
\end{figure*}

\section{Characterization of RuO$_2$ samples by X-ray diffraction, resistivity and magnetic susceptibility measurements}
\label{sec_char}

Figure \ref{Charac}(a) shows the Rietveld refinement against X-ray diffraction powder data for a RuO$_2$ sample obtained by crushing some crystals. 
No impurity peaks could be identified and an undistorted rutile structure describes these data perfectly. The lattice parameters are $a$=4.49109(1)\,\AA{} and $c$=3.10700(1)\,\AA{} in the same range as the ones reported earlier \cite{Berlijn.2017,Rogers.1969}. In addition, neutron Laue images were taken with incoming beam along the [100] direction, showing the perfect crystallinity of our samples (Figure \ref{Charac}(c)). 

Figure \ref{Charac}(b) shows the normalized resistivity of \RuO measured along the [001] and [100] directions with a value of $124\, \Omega$cm at 290\, K. It is nearly identical for the two sample orientations which is in agreement with earlier reports about the isotropic resistivity in \RuO single crystals \cite{Ryden.1970} and thin films \cite{Feng.2022}. The temperature dependence of the resistivity reveals a residual resistivity ratio of 158, which is in agreement with the large values reported in reference \cite{Feng.2022,Ryden.1970,Schafer.1963} and confirms the good sample quality. 

The susceptibility measurements (see figure \ref{Charac}(d)) parallel and perpendicular to the tetragonal axis exhibit no evidence for a magnetic phase transition. 
There is a small magnetic anisotropy with a $\sim$0.1$\times$10$^{-4}$\,emu/mol higher susceptibility along [001]. The temperature dependence is small but quite peculiar. Upon cooling from 300\,K the susceptibility linearly decreases until a minimum near 30\,K followed by an uptake at lower temperatures.

The low temperature values of $1.7\times 10^{-4}$ and  $1.65\times 10^{-4}$\,emu/mol along [001] and [100], respectively,  are in agreement with earlier studies \cite{Ryden.1970,Berlijn.2017}.
Also the temperature dependencies agree with reference \cite{Berlijn.2017}, suggesting that the low temperature enhancement is not entirely due to impurities. 
Figure \ref{Charac}(e) presents the magnetisation determined as a function of applied field at 2\,K yielding a moment of $\sim$0.002\,$\mu_B$/Ru-atom at 7\,T. The magnetization curve is almost perfectly linear, while some impurity moments should follow a non-linear Brillouin function at 2\,K. To qualitatively illustrate this, we present the difference between the linear fit and the data in figure \ref{Charac} (e), which indicates only a tiny impurity contribution. 

The magnetic susceptibility can be compared to that of metallic and nonmagnetic Sr$_2$RuO$_4$, which amounts to $1\times 10^{-3}$\,emu/mol after correcting for the closed shells diamagnetic contributions \cite{Maeno.1997}. The same correction enhances the RuO$_2$ susceptibilities by 0.42$\times$10$^{-4}$\,emu/mol so that the intrinsic paramagnetic susceptibility
is about a factor 5 smaller in RuO$_2$, resulting in the tiny magnetic polarization that can be induced by an external magnetic field. Note that the magnetic
susceptibilities in ferromagnetic or nearly ferromagnetic ruthenates are much higher than that of Sr$_2$RuO$_4$ \cite{Kunkemoller,Nakatsuji}.
From these magnetization measurements on RuO$_2$ no evidence for magnetic ordering can be deduced.

\section{Crystal structure analyses by X-ray and neutron diffraction}
\label{sec_crystal}

For the structural characterization, comprehensive single-crystal X-ray diffraction experiments were carried out on two samples grown via different transport molecules. 
For these experiments small crystals of volumes 75.3 $\times$10$^{-6}$ and 299.7$\times$10$^{-6}$\,mm$^3$ (for RuO$_2$-(O$_2$) and RuO$_2$-(TeCl$_4$), respectively) were chosen in order to suppress extinction and multiple diffraction issues, as well as to limit absorption. 
With the area detector, large data sets were collected up to a maximum resolution of $(\sin(\Theta)/\lambda)_{{max}}$=1.14\AA$^{-1}$. 
Structural models were refined with the Jana program and
the resulting parameters for these samples at 250\,K and 80\,K are listed in table \ref{refinement}.
The small errors as well as the small $R$ values indicate the good data and refinement quality. 
In spacegroup $P4_2/mnm$ the elements $4_2$ and $n$ yield special extinction rules:
For (0,0,$l$), $l$ must be even, for ($h$,0,$l$), $h+l$ must be even, and for (0,$k$,$l$) $k+l$ must be even. The analysis of our complete and far-reaching data sets does not detect 
reflections violating these rules, thus confirming an undistorted rutile-type structure down to 80\,K. Furthermore, the atomic displacement parameters are not enhanced,
suggesting the absence of a hidden or nearby structural instability.
No significant amount of ruthenium vacancies were found, the samples show nearly perfect stoichiometry.

\begin{figure*}
	\centering
	\includegraphics[width=\textwidth]{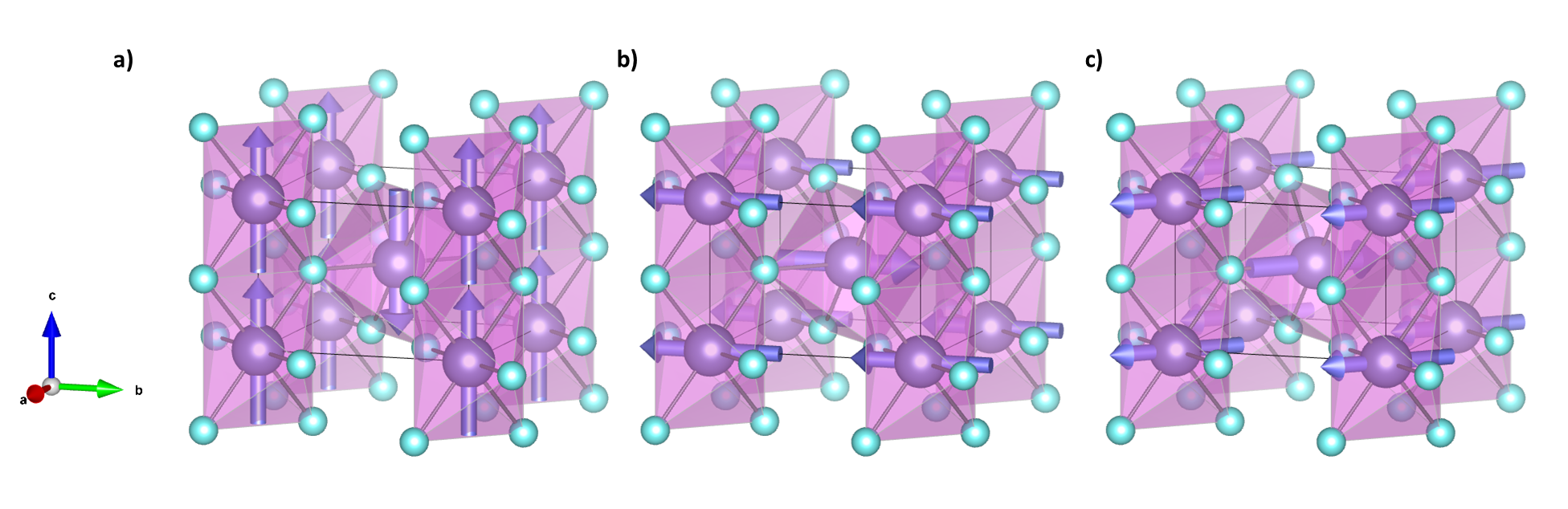}
	\caption{The three possible antiferromagentic structures with the moments (a) along $c$, (b) along $b$ and (c) in plane which could be realised in \RuO. Ruthenium atoms are shown in purple and the oxygen atoms in blue. }
	\label{magnetic_structure}
\end{figure*}

To further strengthen our conclusion about the absence of structural distortions in \RuO , we studied larger crystals with neutron diffraction. On IN12 we used long wavelengths ($k_i$=$2\pi/\lambda$=1.85\,\AA$^{-1}$ and 2.7\,\AA$^{-1}$) and cooled the sample to 1.6\,K.
The main purpose of this IN12 experiment was to search for magnetic scattering but it also allowed us to study nuclear contributions at the (1,0,0), (3,0,0), and (0,0,1) reflections with a strong signal to noise ratio, see figure \ref{Polarisation_analysis}.
In spite of the high statistics there is no evidence for nuclear contribution at
these reflections; any signal must be 1.1$\times$10$^{-3}$ times smaller than that of the weak fundamental reflection (2,0,0). This result sharply contrasts with reference \cite{Berlijn.2017}, which clearly
observes these reflections. We also studied a large volume powder sample on IN12 by scanning 
the 2$\Theta$ ranges of the relevant reflections. A commercial powder
(Thermo Fisher 99.95\% purity) that was annealed at 1000 $^\circ$C to transform the amorphous material into a crystalline one was used. For its characterization see \ref{powApp}. Also this powder experiment cannot detect Bragg reflections at the forbidden $(h,k,l)$ values yielding an upper limit of only 3$\times10^4$ compared to the 1$\times$10$^6$ of the nuclear Bragg reflection (1,1,0) that is weaker than (1,1,1). There is thus no evidence for superstructural reflections violating the ideal rutile-type structure illustrated in figure \ref{PulverIN12} of \ref{powApp}.

A complete data set of integrated Bragg reflection intensities was collected on D9 at 2\,K. For the question of vacancies of the ruthenium site, as well as for the question of a structural distortion, which would include the lighter oxygen atoms, neutrons are better suited. Reflections violating the selection rules of the rutile structure, in particular (1,0,0), (3,0,0), and (0,0,1) are observed in this D9 experiment. 
Since these reflections were absent with much higher statistics in the single crystal IN12 experiment
on the same crystals, they cannot be true Bragg reflections.
We attribute them to multiple diffraction in agreement with reference \cite{Keler.2024}.
Due to the shorter wavelength ($\lambda=0.835$\AA) the probability for multiple diffraction gets strongly
enhanced on D9. Note that the wavelength and the resolution conditions are similar
to those of the experiment by Berlijn $et$ $al.$ \cite{Berlijn.2017}.\\ Multiple diffraction 
arises in the ideal case, when there is a second $(h,k,l)$ vector lying on the Ewald sphere
\cite{Renninger.1937,Zachariasen.1965,Rossmanith.2000}.
If one sets the diffractometer to observe a reflection $(h,k,l)_{\textrm{obs}}$ and if there
is a second reflection $(h,k,l)_{\textrm{sec}}$ that simultaneously lies on the Ewald sphere, the radiation
can be diffracted by this second reflection from $k_i$ to $k'$. For this diffracted 
beam $k'$ the difference reflection $(h,k,l)_{\textrm{dif}}$=$(h,k,l)_{\textrm{obs}}$-$(h,k,l)_{\textrm{sec}}$ lies on
its Ewald sphere so that $k'$ gets diffracted to $k_f$ just as one would expect it for
the initial diffraction process by $(h,k,l)_{\textrm{obs}}$=$(h,k,l)_{\textrm{dif}}$+$(h,k,l)_{\textrm{sec}}$. 
Usually the multiple
diffraction process yields intensities that are small compared to a normal Bragg reflection, but if $(h,k,l)_{\textrm{obs}}$ is a weak or even extinct Bragg reflection the contamination
can become important. In this context, it is worth emphasizing that multiple diffraction cannot violate the translation symmetries yielding the general extinction
rules for (h,k,l) arising from centered lattices. However, the special extinction
conditions arising from non-symmorphic elements as those in RuO$_2$ can be violated by such a process \cite{Braden.1996}.\\ Rotating the sample crystal around its diffraction vector $(h,k,l)_{\textrm{obs}}$ permits one to analyze, whether the reflection is real or contaminated.
This rotation is called a $\Psi$ scan (Renninger scan) and it suppresses a multiple diffraction process while a normal reflection intensity should not depend on $\Psi$. In reality the multiple
diffraction process involves highly indexed Bragg peaks and there are many combinations
so that  $\Psi$ scans reveal many different multiple diffraction events.
The $\Psi$ scans for (1,0,0), (4,0,1) and (3,0,0) shown in figure \ref{Psi} and figure \ref{add_mutiple} of Appendix exhibit intensities
that strongly vary with $\Psi$, revealing the impact of multiple diffraction processes. 
We then performed rocking scans ($\Omega$ scans) for the maximum and minimum intensities of the $\Psi$ scan,
see figure 4 (b)-(d). While scanning at the former $\Psi$ values yields peaks, scans at the latter $\Psi$ values are totally flat unambiguously documenting that these reflections do not have a measurable intrinsic intensity. The same interpretation
holds most likely also for the previous experiment \cite{Berlijn.2017} and was also proposed by another recent neutron based experiment on single crystalline \RuO \cite{Keler.2024}.

The refinement of the rutile structure parameters with the neutron data set taken
at 2\,K yields satisfactory agreement excluding any strong structural distortion
(see table \ref{refinement}). 
The slightly higher $R$-values in comparison to the XRD experiments can be explained by the broad peaks at higher $(h,k,l)$ values as well as by extinction and multiple diffraction problems with the larger crystals required for neutron diffraction. The low-temperature atomic displacement parameters obtained in the neutron experiment at 2\,K do not appear enhanced, as one would expect it in the case of a hidden structural distortion. 
This absence of a structural distortion or of a structural instability is further supported by two DFT calculations of the phonon dispersion \cite{Bohnen.2007,Uchida.2020}, which both do not find unstable phonon modes. 
The neutron refinement yields a small amount of Ru vacancies of 2.6(4)\,\%. Here the
error represents the statistical one from the refinement program which certainly is underestimated as it cannot consider extinction or multiple diffraction issues. Taking also the X-ray results into considerations we estimate our crystals to be nearly stoichiometric with at most a few percent of Ru vacancies.

\begin{figure}
	\centering
	\includegraphics[width=\columnwidth]{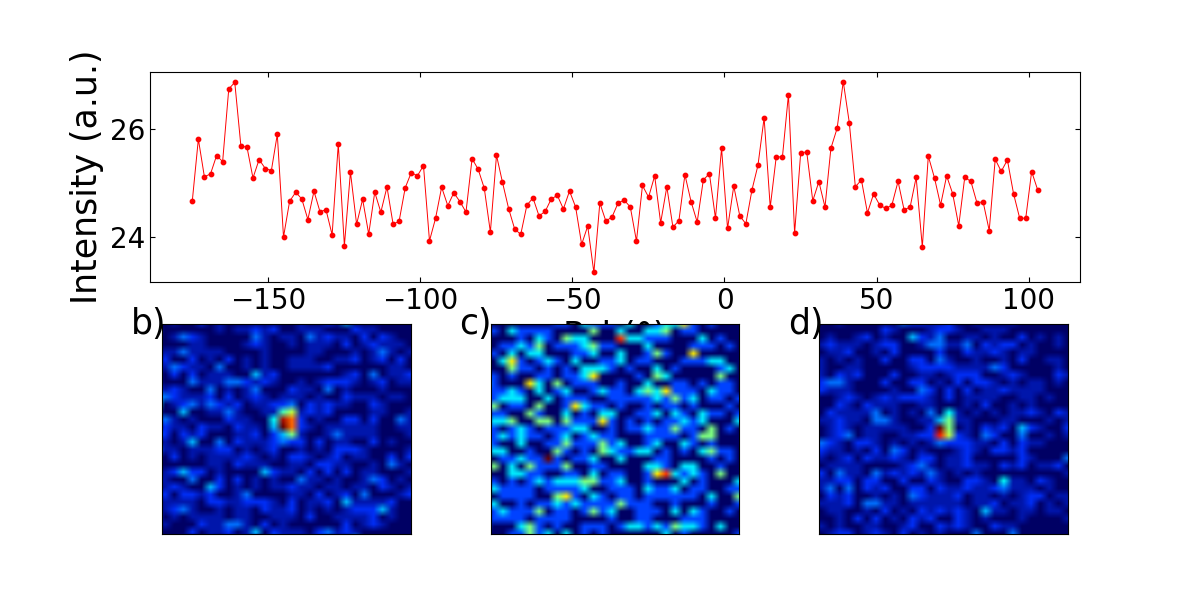}
	\caption{a) $\Psi$ scan of the (100) reflection collected on D9 showing non constant intensity as a clear signature of multiple diffraction. $\omega$ scans at b) $-161\,^\circ$, c) $95\,^\circ$ and d) $10.67\,^\circ$ showing the presence and absence of the reflection for different $\Psi$ values. In the $\omega$ scans with an area detector all intensities are projected along the scan direction. The full quadratic detector frame is shown. }
	\label{Psi}
\end{figure}
\section{Absence of magnetic order}
The main purpose of the IN12 experiment  with incident wavevector $k_i$=1.85\,\AA$^{-1}$ and 2.7\,\AA$^{-1}$ was to study the temperature dependence of the magnetic intensities, in particular at
(1,0,0) and (0,0,1)
Three potential antiferromagnetic structures with a $\vec{q}$=(0,0,0) propagation vector have been proposed for \RuO (figure \ref{magnetic_structure}). These structures would result in a signal at either (1,0,0) or (0,0,1).
Figure \ref{Polarisation_analysis} resumes the polarization analysis for the two measured single crystals at the (1,0,0), (3,0,0) and (0,0,1) reflections. By measuring the three spin-flip
processes for neutron polarization along $\vec{x}$, $\vec{y}$ and $\vec{z}$ directions, one can determine
the anisotropic magnetic signals free of background. We use the conventional coordinate
system in neutron polarization analysis, with $\vec{x}$ along the scattering vector $\vec{Q}$, $\vec{z}$ perpendicular to the scattering plane of the instrument, and $\vec{y}$=$-(\vec{x}\times \vec{z})$.
Neutron scattering only senses the magnetic components perpendicular to $\vec{Q}$. The polarization analysis adds the additional selection rule that a neutron-spin-flip process only involves components perpendicular to the direction of polarization analysis.
Therefore, one can get the magnetic cross sections $\sigma_{x,y,z}$ by simple  substraction. The results are shown in figure \ref{Polarisation_analysis}. Considering the
form factor of Ru$^{4+}$ \cite{Lovesey.2023}, the Lorentz factor and using the nuclear Bragg
peak (2,0,0) for scaling, our results exclude magnetic order of the proposed types with an ordered moment larger than $0.01$\,$\mu_{B}$. Note that the intensity scales
with the square of the moment, so that the proposed value of the ordered moment by Berlijn $et$ $al.$ \cite{Berlijn.2017} would 
be about a factor 25 larger than our detection limit. Our crystals grown with two
different transport molecules do not exhibit sizeable magnetic moment (see \ref{Add}).

Using a much shorter wavelength of 0.835\,\AA, similar to that used by \cite{Berlijn.2017} on the hot diffractometer D3, we obtained a diffraction signal at the critical (h,k,l) positions that must be attributed to multiple diffraction as explained above. 
Performing polarization analysis on a multiple diffraction signal is prone to 
artefacts, in particular one cannot expect to obtain the same flipping ratio as for a fundamental nuclear Bragg peak. Indeed we find different flipping ratios similar to
reference \cite{Berlijn.2017}, but this can not be taken as evidence for magnetic order.

\section{Conclusions}

Single crystals of \RuO grown by two different CTR growth methods (either grown in flowing O$_2$ or with TeCl$_4$ as transporting agent in closed ampoules) are both of high structural quality. Crystals show metallic conductivity behavior and a high 
residual resistivity ratio but no evidence for an anomaly between 300\,K and 4\,K. The 
slightly anisotropic magnetic susceptibility and the magnetization at a field of 7\,T are very small supporting the picture of a Pauli paramagnet. At low temperatures there is a strange enhancement of the susceptibility, but between 30\,K and 300\,K no anomalies
are detected. Comprehensive X-ray and neutron diffraction studies confirm that RuO$_2$ exhibits the ideal rutile structure down to low temperatures. Antiferromagnetic order with a $k$=(0,0,0) propagation vector can be excluded for ordered moments larger than 0.01\,$\mu_B$.

\ack
We acknowledge support by the DFG (German Research Foundation) via Project No. 277146847-CRC 1238 (Subprojects A02 and B04). We thank N.Qureshi for a stimulating discussion.

\begin{appendix}
\section*{Appendix}

\section{Estimation of the theoretical magnetic signal for an ordered moment of $\mathbf{0.05\,\mu_B}$}
\label{Add}

\begin{figure}
	\centering
	\includegraphics[width=\columnwidth]{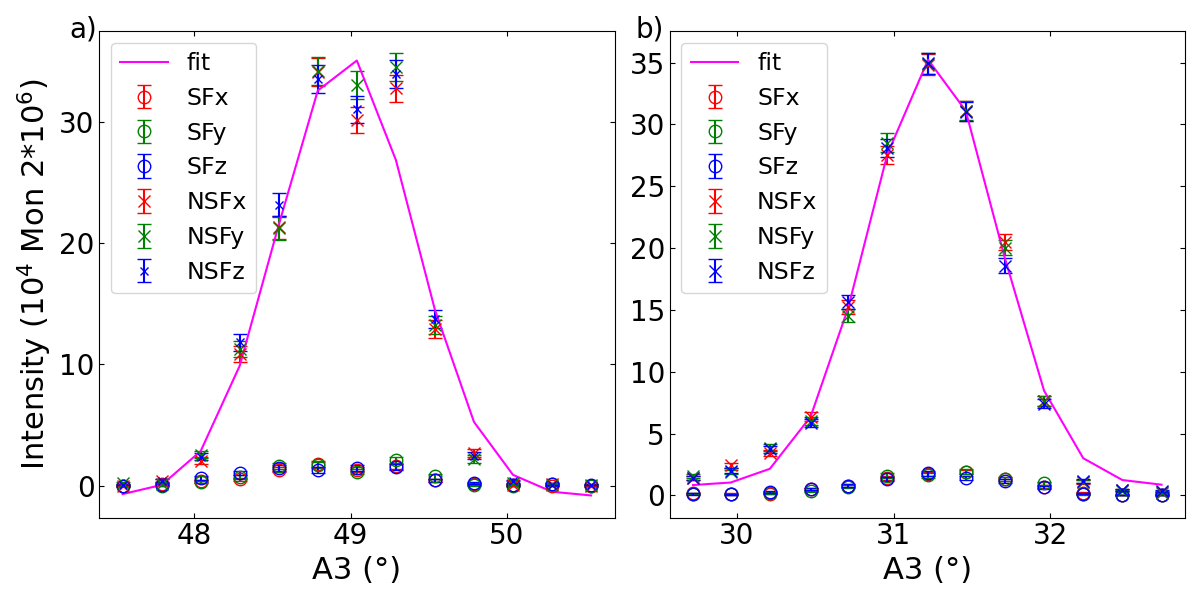}
	\caption{Elastic rocking scans across the (2,0,0) peak for a) the RuO$_2$-(O$_2$) single crystal and b) the RuO$_2$-(TeCl$_4$) crystal measured on IN12, showing no magnetic contribution.}
	\label{Nuclear_peaks}
\end{figure}

For the proposed antiferromagnetic structures in RuO$_2$, see figure \ref{magnetic_structure}, the reflections in the (h0l) plane are either magnetic or nuclear, since the conditions for nuclear and magnetic reflections are h+l=even and h+l=odd, respectively. 
In figure \ref{Nuclear_peaks} the (2,0,0) peak is shown for the two different single-crystal samples measured on IN12. No comparison between the structure factor calculations and the (0,0,2) peak could be made due to strong extinction problems for the strong (0,0,2) reflection.

\begin{figure}
	\centering
	\includegraphics[width=\columnwidth]{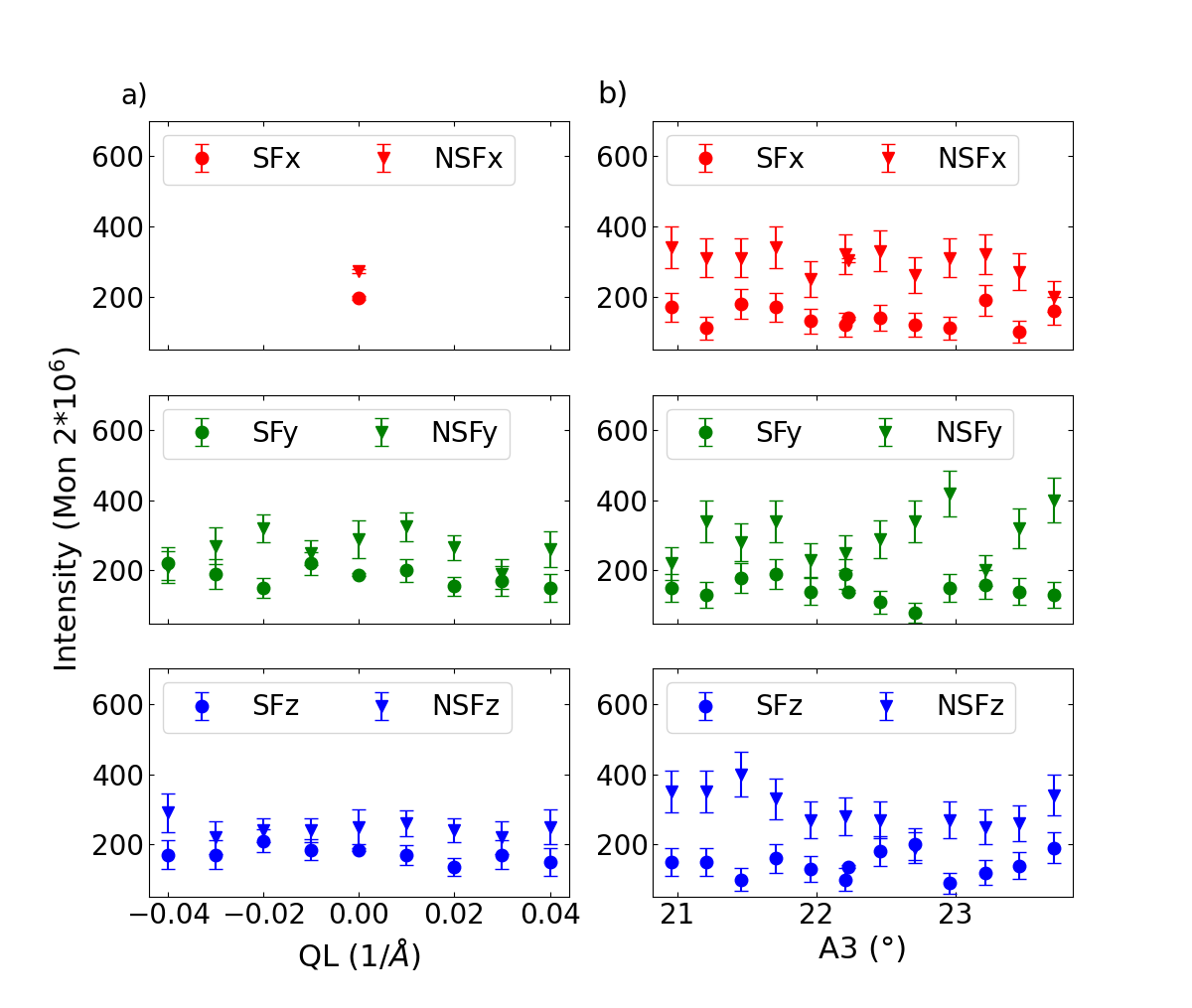}
	\caption{Rocking scans over the potential (1,0,0) magnetic peak for a) RuO$_2$-(O$_2$) and b) RuO$_2$-(TeCl$_4$) single crystal. No signal could be identified in one of the three channels.}
	\label{rocking100}
\end{figure}

The investigation of magnetic elastic scattering in \RuO can be performed by analyzing reflections with $h+l$=odd, specifically the (1,0,0) and (0,0,1). The absence of only one of these reflections would indicate the direction of the magnetic moment.
In order to predict the observable signal within the error bars, we calculate the structure factors.
The potential magnetic intensities were calculated, for the three distinct magnetic models (figure \ref{magnetic_structure}), and for the predicted values of $0.05\,\mu_B$, $0.01\,\mu_B$ and $0.005\,\mu_B$ (table \ref{Calcu}). No signal could be observed in the polarization analysis (figure \ref{Polarisation_analysis}) as well as the rocking scans over the (1,0,0) (figure \ref{rocking100}) and (3,0,0) reflections.

\begin{table}
	\caption{The magnetic signal intensities for the three possible antiferromagnetic structures (M) were estimated by scaling with a nuclear peak. For the single crystals labeled regarding their transporting agent O2 and TeCl$_4$, respectively, the (200) nuclear peak was used as a reference for the intensity. For the powder sample the (101) nuclear peak was used. The calculated intensities are given in counts per Mon $ 2\times 
10^6$. }
\begin{indented}
	\item[]\begin{tabular}{@{}llllll}
		\br
		sam. & M & refl. & $0.05\,\mu_B$& $0.01\,\mu_B$&$0.005\,\mu_B$\\
		\mr	
	RuO$_2$-(O$_2$)& a)& (100) &2006$\pm$211 &80$\pm$8 &20$\pm$2\\
		& b)& (001) &1536$\pm$162 & 61$\pm$6&15$\pm$1\\
		& b)& (100) &1003$\pm$105 & 40$\pm$4&10$\pm$1\\
		& c)& (001) &1536$\pm$162 &61$\pm$6 &15$\pm$1\\
		& c)& (100) &1003$\pm$105 & 40$\pm$4&10$\pm$1\\
		\mr
		RuO$_2$-(TeCl$_4$)& a)& (100) &5552$\pm$238 &222$\pm$10 &55$\pm$2\\
		& b)& (001) &4252$\pm$182 & 170$\pm$7&42$\pm$1\\
		& b)& (100) &2776$\pm$119 & 111$\pm$10&27$\pm$1\\
		& c)& (001) &4252$\pm$182 & 170$\pm$7&42$\pm$1\\
		& c)& (100) &2776$\pm$182 & 111$\pm$10&27$\pm$1\\
		\mr
		pow. & a) & (100) & 664$\pm$10&26.6$\pm$0.4 &6.6$\pm$0.1\\
		 & b) & (001) &508$\pm$8 &20.3$\pm$0.3 &5.08$\pm$0.08\\
		 & b) & (100) &332$\pm$5 &13.3$\pm$0.2 &3.3$\pm$0.05\\
		 & c) & (001) &508$\pm$8 &20.3$\pm$0.3 &5.08$\pm$0.08\\
		 & c) & (100) &332$\pm$5 &13.3$\pm$0.2 &3.3$\pm$0.05\\
		\br
	\end{tabular}
\end{indented}
	\label{Calcu}
\end{table}

\section{Neutron diffraction polarization analysis on a powder sample}
\label{powApp}
\begin{figure}
	\centering
	\includegraphics[width=\columnwidth]{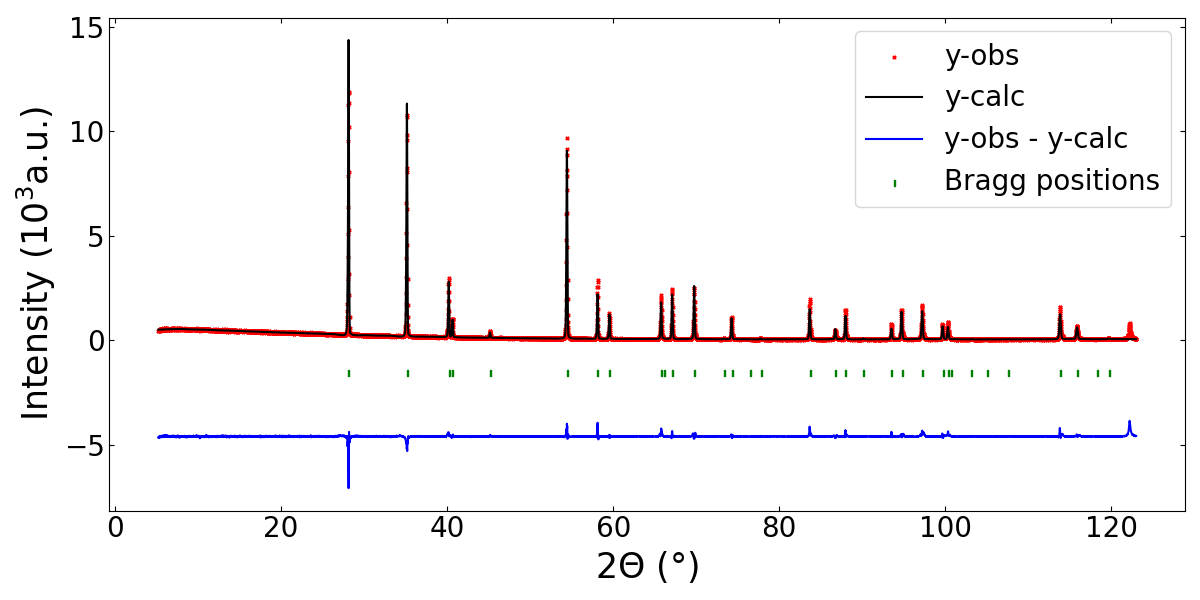}
	\caption{X-ray diffraction data does not show any impurity peaks. Rietveld fit of the \RuO powder synthesized from
amorphous precursor at 1000$^\circ$C gives a $R$-value of 7.47 \%. The lattice parameters are $a$=4.48583(2)\,\AA and $c$=3.10282(1)\,\AA.}
	\label{Rietveld_IN12}
\end{figure}

\begin{figure}
	\centering
	\includegraphics[width=\columnwidth]{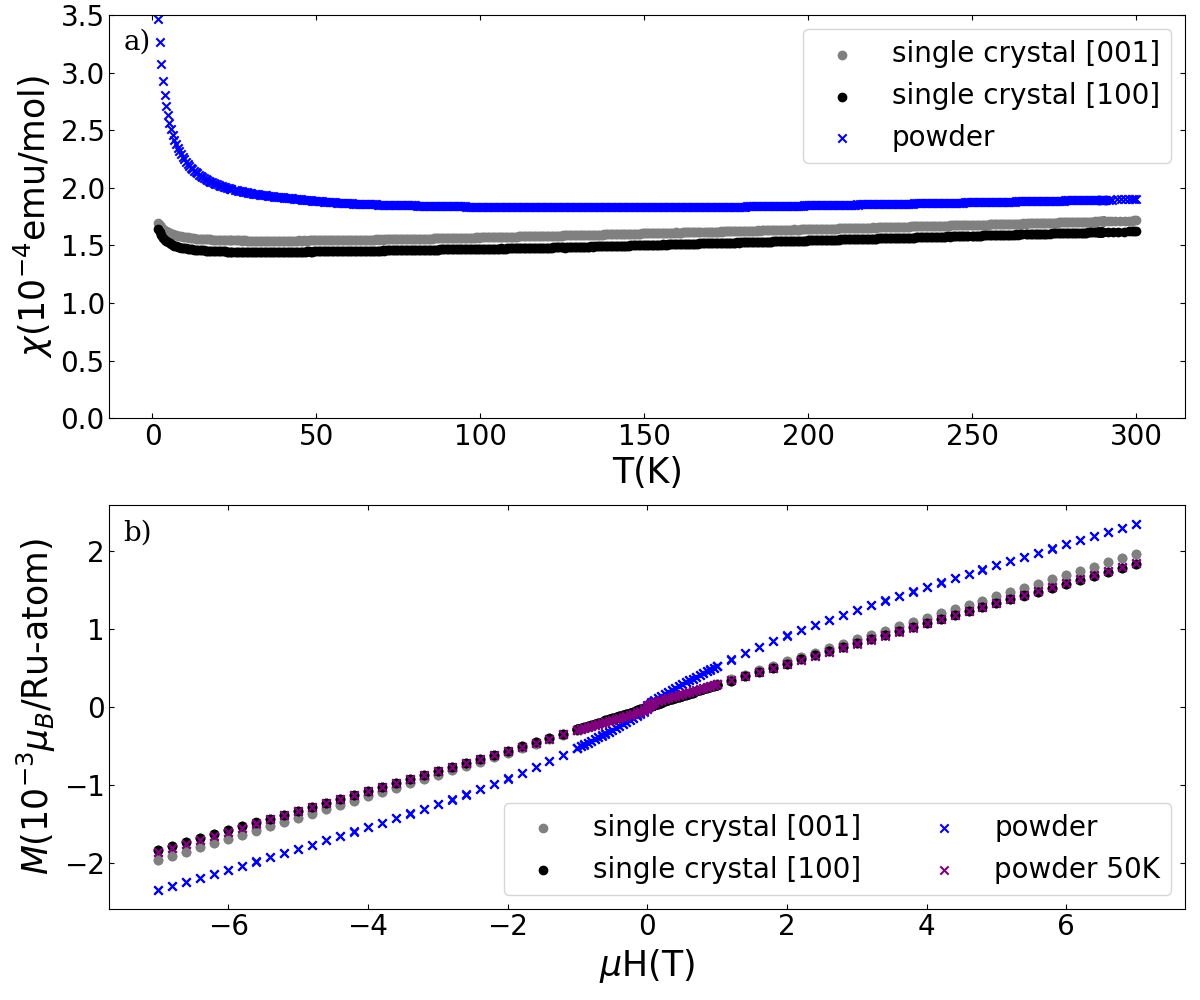}
	\caption{Magnetic characterisation of the powder sample a) temperature dependent susceptibility in comparison to the single crystal (b) Magnetisation for magnetic fields up to 7\,T at 2\,K and 50\,K in comparison to the single crystal data. }
	\label{Mag_pulver}
\end{figure}

Additionally to the two single crystal samples, a powder sample was measured on IN12. The powder was obtained by sintering anhydrous commercial \RuO powder at  $1000\,^{\circ}$C. The quality of the obtained powder sample was ascertained by powder XRD and Rietveld analysis (figure \ref{Rietveld_IN12}). The magnetic susceptibility and magnetisation results for this powder sample are shown in figure \ref{Mag_pulver}. The susceptibility is larger over the entire temperature range compared to the single crystal data and there is a more pronounced uptake at low temperatures which however seems to arise from paramagnetic moments as evidenced by the low-temperature magnetization curve in figure \ref{Mag_pulver} b). The intrinsic magnetic response is only very little enhanced compared to the single crystals.

\begin{figure}
	\centering
	\includegraphics[width=\columnwidth]{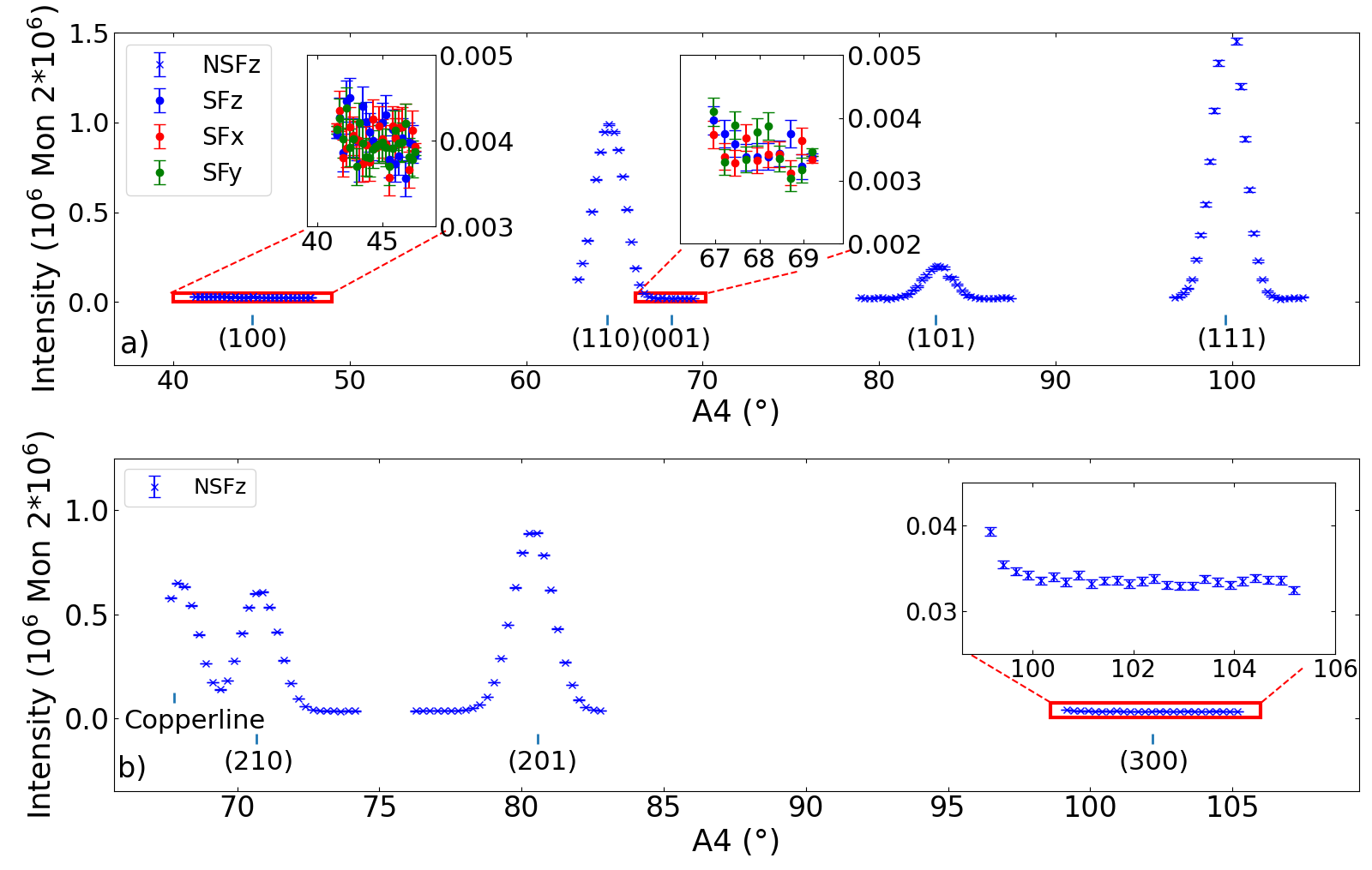}
	\caption{Powder pattern for a) k$_i$=1.85\,\AA${^{-1}}$ and b) k$_i$=2.7\,\AA${^{-1}}$ obtained with IN12 does not show a magnetic signal for the (1,0,0) or (0,0,1) reflections.}
	\label{PulverIN12}
\end{figure}

The neutron diffraction pattern for $k_i$=1.85\,\AA${^{-1}}$ and 2.71\,\AA${^{-1}}$ does not show a magnetic signal of at least 0.04 $\mu_B$ at the (1,0,0), (3,0,0) or (0,0,1) Bragg reflection, see figure \ref{PulverIN12} and \ref{Calcu}. Thus the magnetic order proposed in \cite{Berlijn.2017} would be visible in this powder experiment.

\section{Multiple diffraction in \RuO }
\begin{figure}
	\centering
	\includegraphics[width=\columnwidth]{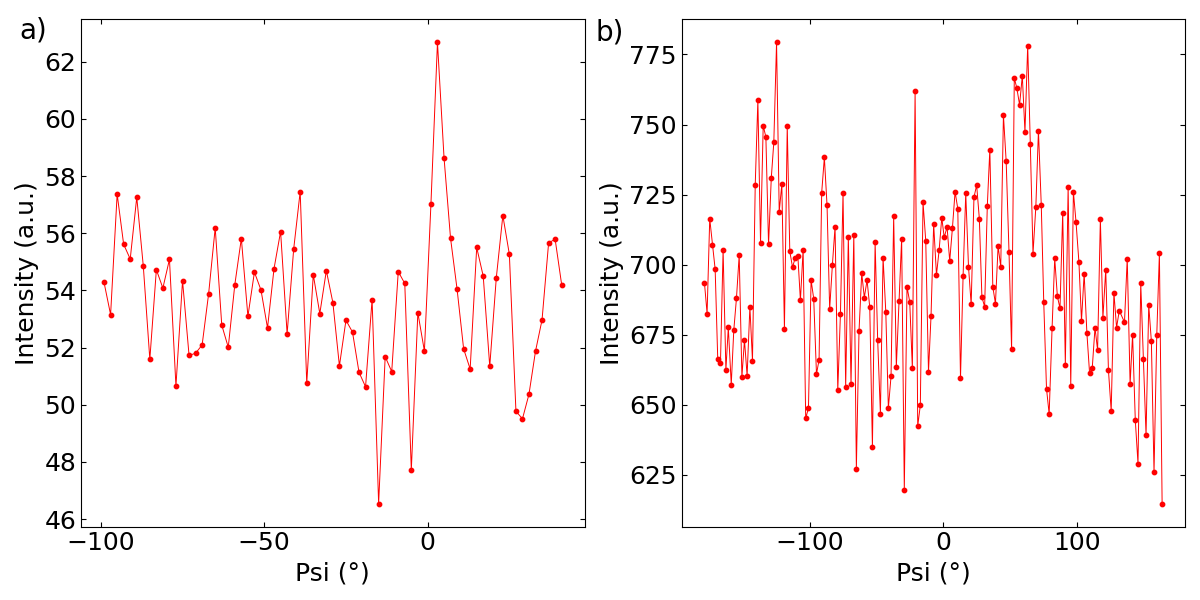}
	\caption{$\Psi$ scans along reflections (a) (401) and (b) (300), with multiple diffraction origin. These reflections have been measured on two different crystals.}
	\label{add_mutiple}
\end{figure}

The experiment on D9 found 74 reflections violating the rutile structure. The origin of all the reflections can be attributed to multiple diffraction in accordance with the absence of such reflections in the IN12 experiment. On D9 the multiple diffraction origin has been verified by $\Psi$-scans, (see figure \ref{Psi} and \ref{add_mutiple}) for three different reflections. 
\end{appendix}
\section*{ORCID iDs}
L. Kiefer\orcidlink{0009-0008-5716-2816}\\
F. Wirth\orcidlink{0000-0002-6386-029X}\\
A. Bertin\orcidlink{0000-0001-5789-3178}\\
P. Becker\orcidlink{0000-0003-4784-3729}\\
L.~Bohat\'{y}\orcidlink{0000-0002-9565-8950}\\
K. Schmalzl\orcidlink{0000-0003-4836-5642}\\
M.~Braden\orcidlink{0000-0002-9284-6585}
\section*{References}

\bibliography{RuO2_4_corrected}

\end{document}